\documentclass[pra,aps,showpacs,twocolumn,eqsecnum,superscriptaddress]{revtex4}
\usepackage{graphicx}
\usepackage{amsbsy}
\usepackage{amsmath,amsthm}
\usepackage{amsfonts}

\newcommand{\nn}{\nonumber \\}
\newcommand{\ex}[1]{\langle{#1}\rangle}
\newcommand{\CS}{\Xi}
\newcommand{\dg}{^\dagger}
\newcommand{\re}{{\rm Re}}
\newcommand{\im}{{\rm Im}}
\newcommand{\A}{\Omega}
\newcommand{\N}{{\cal N}}

\begin{document}

\title{Adaptive phase measurements for narrowband squeezed beams}

\author{Dominic W.\ Berry}
\affiliation{Department of Physics, The University of Queensland, Brisbane,
Queensland 4072, Australia} \email{berry@physics.uq.edu.au}
\author{Howard M.\ Wiseman}
\affiliation{Centre for Quantum Dynamics, School of Science, Griffith
University, Nathan, Brisbane, Queensland 4111, Australia}
\email{H.Wiseman@griffith.edu.au}
\date{\today}

\begin{abstract}
We have previously [Phys.\ Rev.\ A {\bf 65}, 043803 (2002)] analyzed adaptive
measurements for estimating the continuously varying phase of a coherent beam,
and a broadband squeezed beam. A real squeezed beam must have finite photon
flux ${\cal N}$ and hence can be significantly squeezed only over a limited
frequency range. In this paper we analyze adaptive phase measurements of this
type for a realistic model of a squeezed beam. We show that, provided it is
possible to suitably choose the parameters of the beam, a mean-square phase
uncertainty scaling as $({\cal N}/\kappa)^{-5/8}$ is possible, where $\kappa$
is the linewidth of the beam resulting from the fluctuating phase. This is an
improvement over the $({\cal N}/\kappa)^{-1/2}$ scaling found previously for
coherent beams. In the experimentally realistic case where there is a limit on
the maximum squeezing possible, the variance will be reduced below that for
coherent beams, though the scaling is unchanged.
\end{abstract}
\maketitle

\section{Introduction}
\label{sec:intro}
Optical phase measurements are a valuable means for high precision measurement
of displacement, for example for gravitational wave detection \cite{grav}. Two
types of phase measurement can be distinguished. The first is measuring the
phase of a beam or pulse of light relative to that of a strong local oscillator,
which is treated classically. The second is measuring the phase shift between
the light beams in the two arms of an interferometer, both of which are treated
quantum mechanically. In this paper we are concerned solely with the first type.

Phase measurements are most easily analyzed, and thus best understood, for a
single mode pulse with a fixed phase. The optimal measurement scheme (for all
commonly produced states) for such a single-shot measurement is the so-called
canonical measurement \cite{canon}. For coherent states, where the quantum noise
is independent of the quadrature, the canonical phase variance \cite{canon}
asymptotes $1/4\bar n$ in the large $\bar{n}$ limit, where $\bar n$ is the mean
photon number. The $1/\bar{n}$ scaling is referred to as  the standard quantum
limit (SQL). On the other hand, for more general states where the noise is
quadrature-dependent, the canonical phase variance may approach the Heisenberg
limit $1.89/\bar n^2$ \cite{heis}.

Unfortunately it is not possible to achieve canonical measurements with linear
optical elements (unless one discards most of the measurement results
\cite{preg}). One solution that has been developed is to use feedback. The field
is combined with a local oscillator at a beam splitter, and partial results
during the measurement are fed back to adjust the phase of the local oscillator
for measuring the field in the next part of the pulse. This approach has been
extensively studied for single-shot measurements, and it has been shown that it
can achieve scalings almost at the Heisenberg limit \cite{single}. It has also
been experimentally realized \cite{mabuchi}, verifying an improvement over
non-adaptive (heterodyne) detection. However, because the experiment used
coherent states, it was not possible to verify a scaling better than the SQL.

In practice it is easier to produce a continuous squeezed beam rather than a
squeezed pulse of light. This motivates considering continuous, rather than
single-shot, measurements. In the continuous case, if the phase to be measured
is constant, then the variance will become arbitrarily small with time. To
obtain a nontrivial result, it is necessary to consider a phase which varies in
time. We wish to determine how accurately the measurement scheme estimates this
varying phase. The simplest model for a varying phase is a Wiener process; that
is, its rate of change is Gaussian white noise with intensity $\kappa$, giving
rise to a Lorentzian lineshape of the beam with width $\kappa$.

Continuous measurements of this type were considered in Refs.\ \cite{cont,pope}.
In Ref.\ \cite{cont}, coherent beams and broadband squeezed beams were
considered, and a simple method for filtering the data to obtain the phase
estimate was used. A scaling law for the optimal variance (or mean-square error)
of the squeezed beam was derived which showed an improvement over the coherent
beam result. However, as was pointed out in Ref.~\cite{pope}, this analysis had
two short-comings (which tend to counteract each-other). First, it
considered only the photon flux due to the coherent component of the beam,
whereas in fact the photon flux from the broadband squeezing is strictly
infinite. Second, its filtering technique ignored the fact that phase
information may be obtained from the photocurrent {\em noise} \cite{noise}.
For broadband squeezing the information from the noise is strictly infinite,
and would allow the phase to be determined exactly, modulo $\pi$.

Ref.\ \cite{pope} adopted a more sophisticated (Bayesian) filtering of the
state, but restricted its attention to coherent states.  It found different
results for $\N \alt \kappa$, though the asymptotic variance of
$(1/2)\sqrt{\kappa/\N}$ for $\N \gg \kappa$ was the same as found in
Ref.~\cite{cont}. An asymptotic improvement by a factor of $1/\sqrt{2}$ over
non-adaptive (heterodyne) detection was also confirmed.

In this paper we apply a Bayesian approach to phase measurements on a
{\em narrowband} squeezed beam. This estimation procedure optimally uses the
phase information obtained from both the mean field and from the noise. The
narrowband squeezing ensures that the photon flux, and the rate of information
accumulation in the photocurrent record, are finite. To obtain analytical
results, we concentrate on the asymptotic regime. We find an approximate
analytical expression for the optimal scaling of the phase variance as
$(\N/\kappa)^{5/8}$, and confirm this via numerical simulations. Our analysis
confirms that there is a scaling advantage over the SQL, although it is slightly
less than we had previously thought \cite{cont}.

The remainder of this paper is organized as follows. In Sec.\ \ref{simple} we
give a simple explanation of the scaling, and explain why it is different from
that in Ref.\ \cite{cont}. Then we give the Bayesian analysis of measurements
with feedback in Sec.\ \ref{sec:bay}, and of heterodyne measurements in Sec.\
\ref{sec:het}. We revisit the scaling based on the Bayesian treatment in Sec.\
\ref{sec:rev}, then give numerical results in Sec.\ \ref{sec:num}. We conclude
in Sec.\ \ref{sec:conc}, and give additional details of the derivations in the
appendices.

\section{Simple explanation of scaling}
\label{simple}
In this section we give a simple explanation to predict the scaling of the
phase variance under continuous adaptive measurements. First we discuss the
simple case of a coherent beam, then we proceed to the case of squeezing.

\subsection{Coherent states}
\begin{figure}
\centering
\includegraphics[width=0.35\textwidth]{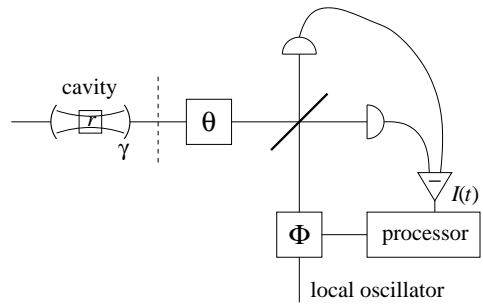}
\caption{The model of the experiment. The phase shift $\theta$ is imposed on the
continuous beam, and a phase shift $\Phi$ is imposed on the local oscillator.
These beams are incident on a 50:50 beam splitter, and the difference
photocurrent $I(t)$ is determined. The processor then adjusts $\Phi$ based on
$I(t)$.  The simplest case is when the signal beam is in a coherent state, but
for more accurate phase estimation a squeezed beam, produced by the apparatus to
the left of the dashed vertical line, is used. This consists of a parametric
down-converter characterized by intensity damping rate $\gamma$ and $\chi^{(2)}$
nonlinearity parameterized by $r$.} \label{fig0}
\end{figure}

The configuration considered is as in Fig.\ \ref{fig0}, although for the moment
everything to the left of the dashed line should be ignored. The coherent signal
beam  has photon flux (i.e.\ mean number of photons per unit time) $\N$. A phase
shift $\theta(t)$ (to be estimated by the experimenter) is imposed on the
signal, and a known phase shift of $\Phi(t)$ may be imposed by the experimenter
on the local oscillator. These are combined at a beam splitter, and the
difference photocurrent $I(t)$ gives a measurement of a quadrature of the
signal.  Feedback may be used to determine the phase $\Phi(t)$ for the local
oscillator using a signal processor to filter the signal $I(t)$.  The estimate
of $\theta(t)$  is also determined based on $I(t)$.

For fixed system phase and a coherent beam, the variance
for adaptive measurements over time interval $\Delta t$ is
\begin{equation}
\label{adcoh}
\sigma^2 \approx \frac {1}{4\Delta t\N}.
\end{equation}
This is simply the standard result for adaptive measurements on a coherent field
with a mean photon number of $\Delta t\N$ \cite{single}. If there is an estimate
of the phase at time $t$, then the phase estimate at time $t+\Delta t$ may be
taken to be a weighted average of the estimate at time $t$ and the estimate
obtained from data in the time interval $\Delta t$. The variance at time
$t+\Delta t$ will then satisfy
\begin{equation}
\label{invinc}
\sigma_{t+\Delta t}^2 \approx \frac 1{1/{\sigma_t^2}+ 4\Delta t\N}.
\end{equation}

Now we assume that the phase fluctuations on the signal are white:
\begin{equation}
\dot\theta = \sqrt{\kappa}\zeta,
\end{equation}
where $\zeta$ is Gaussian white noise satisfying $\langle \zeta(t)\zeta(t')
\rangle=\delta(t-t')$. Taking account of these fluctuations, we expect the
variance at time $t+\Delta t$ to be
\begin{equation}
\sigma_{t+\Delta t}^2 \approx \frac 1{1/{\sigma_t^2}+ 4\Delta t\N}+
\kappa\Delta t.
\end{equation}
Provided $\kappa\Delta t \ll \sigma_t^2 \ll 1/(4\Delta t \N)$,
the total change in the inverse variance is
\begin{equation}
\label{cohsted}
\Delta\left(\frac 1{\sigma^2}\right) \approx -\Delta t\frac{\kappa}
{\sigma^4} + 4\Delta t \N.
\end{equation}
The steady-state variance is therefore
\begin{equation}
\sigma^2 \approx \frac 12 \sqrt{\kappa/\N}.
\end{equation}
This is just what was obtained in Ref.\ \cite{cont}.

In the case of heterodyne measurements, the variance for measurements over time
interval $\Delta t$ is twice that in Eq.\ \eqref{adcoh} \cite{single}. Then the
last term in Eq.\ \eqref{cohsted} is $2\Delta t\N$, and the steady state
variance is
\begin{equation}
\label{het}
\sigma^2 \approx \frac 1{\sqrt{2}} \sqrt{\kappa/\N}.
\end{equation}
Thus the adaptive measurements give a $1/\sqrt{2}$ reduction in the variance
over heterodyne measurements.

\subsection{Squeezed states}
\label{sec:sqst}
We may use a similar method in the case of squeezed states. In this case we have
a number of new features to the model, indicated in Fig.~1 by the apparatus to
the left of the dashed line. The squeezed beam is produced by a cavity with
decay constant $\gamma$.  The beam has squeezing parameter $r$ and coherent
amplitude $E$. The flux is given by \cite{GarZol00}
\begin{equation}
\label{flux}
\N = \frac{E^2}4 + \frac{\gamma}2\sinh^2 r.
\end{equation}

For squeezing which is not too large, the phase variance for adaptive
measurements over time interval $\Delta t$ is, using the simple estimation
technique of Ref.~\cite{cont},
\begin{equation}
\sigma^2 \approx \frac{1}{e^{2r}\Delta t E^2}.
\end{equation}
That is, the variance is reduced by a factor of $e^{2r}$ from what would be
obtained for a coherent state. The equivalent of Eq.\ \eqref{cohsted} is then
\begin{equation}
\label{steady}
\Delta\left(\frac 1{\sigma^2}\right) \approx -\Delta t\frac{\kappa}
{\sigma^4} + e^{2r}\Delta t E^2.
\end{equation}
Provided the dominant contribution to the flux is the coherent component, the
steady-state variance is
\begin{equation}
\label{sig}
\sigma^2 \approx \frac 12 e^{-r}\sqrt{\kappa/\N}.
\end{equation}

There is a limit on how large the squeezing can be before this approximation
fails. In Ref.\ \cite{cont} the limitation considered was the error in the
feedback phase, which causes the measurement to not be exactly on the squeezed
quadrature. This limitation gives an overall scaling of $(\kappa/\N)^{2/3}$.

Here we consider the additional limitation due to the finite squeezing
bandwidth. For measurements performed over a time scale shorter than $1/\gamma$,
no squeezing will be observed. In fact, it is necessary to perform measurements
over a time scale of order $e^r/\gamma$ before squeezing is observed. For $e^r$
large, this time is approximately the reciprocal of the decay constant for one
of the cavity quadratures [see Eq.\ \eqref{anty} in the following section].
However, it is not obvious why this decay constant is important, because it
relates to the antisqueezed quadrature. We give a more thorough explanation of
this time scale in the following section.

The phase information used to obtain the phase estimate will be from a
finite time interval. For times before $t-\sigma^2/\kappa$, the system phase
will differ from the current system phase by an amount comparable with the phase
uncertainty. Therefore, the majority of the phase information must be taken from
a time interval of length approximately $\sigma^2/\kappa$. We use $\chi$ to
denote the inverse of this time interval, so $\chi\sim\kappa/\sigma^2$.

In order to observe squeezing, this time interval must be longer than the time
scale $e^r/\gamma$, so we require $\sigma^2/\kappa>e^r/\gamma$. Here we are
concerned with scaling, so we ignore multiplicative constants in the rest of
this section. In order to obtain the scaling we also need to consider the
limitation due to the squeezing contribution to the flux.  From Eq.\
\eqref{flux}, we have $\N>\gamma e^{2r}$, so
\begin{equation}
\label{lower}
\sigma^2 > \frac\kappa\N e^{3r}.
\end{equation}
In order to obtain the minimum phase variance, we need to take the squeezing $r$
to be as large as possible consistent with the Eqs.~(\ref{sig}) and
(\ref{lower}). That gives $e^r\sim(\N/\kappa)^{1/8}$, so
\begin{equation}
\label{scaling}
\sigma^2 \sim \left( \frac\kappa\N\right)^{5/8}.
\end{equation}

Note that if we had used the time scale $1/\gamma$, rather than $e^r/\gamma$,
then we would have obtained the inequality $\sigma^2/\kappa>1/\gamma$. Then
Eq.\ \eqref{lower} would become $\sigma^2 > e^{2r}\kappa/\N $, and the maximum
squeezing would be $e^r\sim(\N/\kappa)^{1/6}$. This then would give the variance
scaling as $\sigma^2 \sim \left( \kappa/\N\right)^{2/3}$, which is identical to
that in Ref.\ \cite{cont}. Thus we can see that the extra $e^r$ is essential to
obtaining the different scaling here. This scaling will be shown to be correct
numerically in Sec.\ \ref{sec:num}.

In practice it is not possible to achieve arbitrary squeezing; typically the
maximum value of $e^{2r}$ achieved is about 2 \cite{perscomm}. In that case, we
simply have the variance given in Eq.\ \eqref{sig}, with $r$ taken to be the
maximum experimentally achievable value. That is, the scaling is the same as for
coherent states, but the variance is reduced by a factor of $e^{-r}$.

Note that we have the limitations on $\gamma$ (ignoring constant factors)
\begin{equation} \label{14}
e^{2r} \sqrt{\frac\N \kappa} < \frac \gamma\kappa < e^{-2r} \frac \N \kappa.
\end{equation}
These limitations may be satisfied provided $\N/\kappa > e^{8r}$; for
$e^{2r}\sim 2$ this limit is about 16. Experimentally there is only limited
control over the value of $\gamma$. However, for larger $\N/\kappa$ there is a
wide range of values for which a reduction in the phase variance should be
observed.

\section{Adaptive measurements}
\label{sec:bay}
Now we give the detailed Bayesian analysis of the phase estimates for narrowband
squeezing. We will use this to justify some of the steps used in the previous
section, as well as to estimate the additional phase information that may be
obtained from the noise. A continuous squeezed beam produced by a cavity may be
modeled by the operator equations in the Heisenberg picture \cite{drummond}
\begin{align}
\frac{d\hat x}{dt} &= -\hat x \gamma(1+\varepsilon)/2 +\sqrt{\gamma}\hat\xi \\
\label{anty}
\frac{d\hat y}{dt} &= -\hat y \gamma(1-\varepsilon)/2 +\sqrt{\gamma}\hat\eta \\
\hat I &= \cos(\Phi-\theta)[\sqrt{\gamma}\hat x-\hat \xi]
+\sin(\Phi-\theta)[\sqrt{\gamma}\hat y+E-\hat \eta].
\end{align}
These equations are equivalent to Eqs.\ (4.48) and (4.54) in
Ref.\ \cite{drummond}. The quantities $\hat x$ and $\hat y$ are the two
quadratures of the cavity field, $\hat\xi$ and $\hat\eta$ are the quadrature
noise operators, $\gamma$ is the cavity decay constant (equivalent to $\kappa_1$
in \cite{drummond}), and $\varepsilon$ is a constant related to the usual
squeezing parameter by
\begin{equation}
e^r = \frac{1+\varepsilon}{1-\varepsilon}.
\end{equation}
$\hat I$ is the output quadrature at angle $\Phi-\theta$; the photocurrent
measured corresponds to the measured value of this operator. Note that the
squeezed quadrature here is $x$ (rather than $y$, as in \cite{drummond}). Also
we have added the displacement $E$, so we obtain a squeezed coherent field
rather than a squeezed vacuum.

These equations may be solved by using the Wigner distribution, and replacing
the quadrature operators with the corresponding quadrature variables for the
Wigner distribution. Also the output quadrature $\hat I$ is replaced with
the detected photocurrent $I$.
\begin{align}
\label{system1}
\dot x &= -x \gamma(1+\varepsilon)/2 +\sqrt{\gamma}\xi \\
\label{system2}
\dot y &= -y \gamma(1-\varepsilon)/2 +\sqrt{\gamma}\eta \\
I &= \cos(\Phi-\theta)[\sqrt{\gamma}x-\xi]
+\sin(\Phi-\theta)[\sqrt{\gamma}y+E-\eta].
\end{align}
Here $\xi$ and $\eta$ are Gaussian increments satisfying
$\ex{\xi(t)\xi(t')}=\ex{\eta(t)\eta(t')}=\delta(t-t')$.

In order to apply the Bayesian approach, we assume for the moment that both
$\Phi$ and $\theta$ are constant. The results obtained using these assumptions
should also be accurate for cases where these phases only change by a small
amount over the time interval considered. We discretize the equations to give
\begin{align}
\Delta x&=-x \Delta t \gamma(1+\varepsilon)/2+\tilde\xi\sqrt{\gamma\Delta t} \\
\Delta y&=-y \Delta t \gamma(1-\varepsilon)/2+\tilde\eta\sqrt{\gamma\Delta t} \\
I \Delta t &= \{\cos(\Phi-\theta)[\sqrt{\gamma\Delta t}x-\tilde\xi] \nn & \quad
+\sin(\Phi-\theta)[\sqrt{\gamma\Delta t}(y+E')-\tilde\eta]\}\sqrt{\Delta t}.
\end{align}
Here $\sqrt{\gamma}E'=E$, and we have used tildes to indicate that the
stochastic increments have been replaced with Gaussian random variables (with
mean 0 and variance 1).

At all times the experimenter's knowledge can be represented by a probability
distribution for $x$, $y$, and $\theta$ which is Gaussian for $x$ and $y$:
\begin{align}
\label{multi}
P(\vec x,\theta) &= P(\theta)\frac{ \sqrt{\det G(\theta)}}{2\pi}
 e^{-\frac 12[\vec x-\bar x(\theta)]^T G(\theta) [\vec x-\bar x(\theta)]}
 \nn & = P(\theta)P(\vec x|\theta),
\end{align}
where $\vec x=(x,y)^T$ and $\bar x(\theta)$ is the mean of $\vec x$. $P(\theta)$
gives the correct probability for $\theta$ averaging over $x$ and $y$, and the
inverse of $G(\theta)$ is the covariance matrix for $\vec x$ for a given
$\theta$.

\subsection{Phase information}
First we consider the update to the probability distribution due to the
information from the measurement. At each time step we update the probability
distribution using Bayes' rule \cite{pope}:
\begin{equation}
P(\vec x,\theta) \to P(\vec x,\theta)\times P(I|\vec x,\theta).
\end{equation}
Here the constant factor $1/P(I)$ is omitted, because we can normalize at the
end of the calculation. The probability $P(I|\vec x,\theta)$ is given by
\begin{align}
P(I|\vec x,\theta)&=\sqrt{\frac{\Delta t}{2\pi}}
\exp[-\Delta t \{I-\sqrt{\gamma} [\cos(\Phi-\theta)x \nn
&\quad +\sin(\Phi-\theta)(y+E')]\}^2/2] \nn
&=\sqrt{\frac{\Delta t}{2\pi}}
\exp\left[-\frac{\gamma \Delta t}2(\vec x-B)^T A A^T(\vec x-B)\right],
\end{align}
where
\begin{equation}
A = \left[\begin{array}{*{10}c} c \\
s \end{array}\right], \qquad
B = \frac {\Delta I}{2\sqrt\gamma}
\left[\begin{array}{*{10}c}1/c \\
1/s \end{array}\right],
\end{equation}
where $c=\cos(\Phi-\theta)$, $s=\sin(\Phi-\theta)$ and $\Delta I=I-sE$.

We therefore obtain
\begin{align}
&P(I|\vec x,\theta)P(\vec x,\theta)\propto\sqrt{\det G} \nn &\times
\exp\left\{-\frac 12 [(\vec x-\bar x^{(2)})^T G^{(2)}(\vec x-\bar x^{(2)})
\right. \nn &\left.-(\bar x^{(2)})^T G^{(2)} \bar x^{(2)}+\bar x^T G
\bar x+\gamma\Delta t B^T AA^T B] \right\},
\end{align}
where
\begin{align}
G^{(2)}&=G+\gamma \Delta t AA^T \\
\bar x^{(2)} &= (G^{(2)})^{-1} (G\bar x +\gamma\Delta t AA^TB),
\end{align}
Hence the updated probability distribution for $\theta$ is
\begin{align}
P^{(2)}(\theta) &\propto P(\theta)
\sqrt{\frac{\det G}{\det G^{(2)}}}\exp \left\{ -\frac 12 [\bar x^T G \bar x
\right. \nn & \quad \left.+\gamma\Delta t(B^T A)^2-(\bar x^{(2)})^T G^{(2)}
\bar x^{(2)}]\right\}.
\end{align}
In the limit of small $\Delta t$ we obtain the differential equations
\begin{align}
dG&=\gamma dt AA^T, \\
d\bar x &= dt G^{-1}[\sqrt\gamma \Delta IA-\gamma AA^T\bar x], \\
d[\log P(\theta)] &= K -\frac {dt}2 (\Delta I-\sqrt\gamma
A^T \bar x)^2 , \label{35}
\end{align}
where $K$ is a constant; $K$ may be ignored, as it only changes the
normalization.

\subsection{Increments in $x$ and $y$}
Now we take account of the increments in $x$ and $y$. Given the measurement
result $I$, we have the restriction that
\begin{equation}
c\tilde\xi+s\tilde\eta = \sqrt{\Delta t}[\sqrt{\gamma}(cx+sy)-\Delta I].
\end{equation}
We define the new variable
\begin{align}
\tilde\mu & = -(\tilde\xi-c\sqrt{\Delta t}[\sqrt{\gamma}(cx+sy)-\Delta I])/s \nn
& = (\tilde\eta-s\sqrt{\Delta t}[\sqrt{\gamma}(cx+sy)-\Delta I])/c.
\end{align}
This is simply a Gaussian random variable with mean zero and variance 1.
In terms of this the new difference equations are
\begin{align}
\Delta x &= -x \Delta t \gamma(1+\varepsilon)/2 -c\Delta t[\sqrt{\gamma}\Delta I
-\gamma(cx+sy)] \nn & \quad -s\tilde \mu\sqrt{\gamma\Delta t} \\
\Delta y &= -y \Delta t \gamma(1-\varepsilon)/2 -s\Delta t[\sqrt{\gamma}\Delta I
-\gamma(cx+sy)] \nn & \quad +c\tilde \mu\sqrt{\gamma\Delta t}.
\end{align}

The deterministic part of the increment gives the mapping
$\vec x \mapsto C\vec x-\sqrt\gamma\Delta t \Delta I A$ with
\begin{align}
C &= \openone - \Delta t \gamma[( \openone+\varepsilon\Sigma)/2- AA^T],\nn
\Sigma &= \left[\begin{array}{*{10}cc}
1 & 0 \\ 0 & -1 \\ \end{array}\right].
\end{align}
The stochastic part increases the covariance matrix for $\vec x$ according to
$G \mapsto (G^{-1}+\gamma\Delta t\Pi)^{-1}$, where
\begin{equation}
\Pi = \left[\begin{array}{*{10}cc}
s^2 & -sc \\ -sc & c^2 \\ \end{array}\right].
\end{equation}
Overall, we update the covariance matrix and mean to
\begin{align}
G^{(3)} &= [C(G^{(2)})^{-1}C+\gamma\Delta t \Pi]^{-1}, \nn
\bar x^{(3)} &= C\bar x^{(2)}-\sqrt\gamma\Delta t \Delta I A.
\end{align}

In the limit of infinitesimal $\Delta t$ we again obtain differentials, which
when added to those at the end of Sec.~III~A give
\begin{align}
\label{gdif}
dG &= \gamma dt[ AA^T + G +\varepsilon (\Sigma G+G\Sigma)/2 \nn &
\quad - (AA^T G + G AA^T +G\Pi G)], \\
\label{xdif}
d\bar x &= -\gamma dt \, (\openone+\varepsilon\Sigma)\bar x/2 \nn & \quad
+\sqrt\gamma dt (G^{-1}-\openone)A(\Delta I-\sqrt\gamma
A^T \bar x).
\end{align}

\subsection{Solution}
From Eq.~\eqref{35}, the probability distribution for the phase is obtained by
integrating over
$(\Delta I-\sqrt\gamma A^T\bar x)^2$. From the reasoning given in Appendix
\ref{ap:var}, we may estimate the final phase variance by determining the
expectation value for a given phase $\Theta$, and expanding to second order in
$\theta-\Theta$. The probability distribution for $I$ is determined based on
$\Theta$, and we determine the probability distribution for $\theta$ based on
the measurement result $I$. We interpret $\Theta$ as the actual system phase,
and $\theta$ as a dummy variable used for the probability distribution.

For fixed system phase and feedback phase, $G$ will reach an equilibrium value.
To estimate this equilibrium value, we rotate $\vec x$ via
\begin{equation}
R=\left[ \begin{array}{*{10}cc}
c & s \\ -s & c \\ \end{array}\right],
\end{equation}
which is defined to simplify $AA^T$ and $\Pi$.
Using bars to denote the rotated values of variables, so for example
$\bar G^{-1}=RG^{-1}R\dg$, we have from Eq.~\eqref{gdif} at steady state
\begin{align}
&\bar G^{-1}\left[ \begin{array}{*{10}cc}1 & 0 \\ 0 & 0 \\ \end{array}\right]
\bar G^{-1}+\bar G^{-1} +\varepsilon (\bar\Sigma \bar G^{-1}
+\bar G^{-1}\bar\Sigma)/2 \nn &=\left[ \begin{array}{*{10}cc}
1 & 0 \\ 0 & 0 \\ \end{array}\right]\bar G^{-1} +\bar G^{-1}
\left[ \begin{array}{*{10}cc}
1 & 0 \\ 0 & 0 \\ \end{array}\right]
+\left[ \begin{array}{*{10}cc} 0 & 0 \\ 0 & 1 \\ \end{array}\right],
\end{align}
with
\begin{equation}
\bar\Sigma = \left[ \begin{array}{*{10}cc}
c^2-s^2 & -2sc \\ -2sc & s^2-c^2 \\ \end{array}\right].
\end{equation}

Using the notation $a=(\bar G^{-1})_{00}$, $b=(\bar G^{-1})_{01}$,
$d=(\bar G^{-1})_{11}$, $X=(1-\varepsilon)/2+s^2\varepsilon$,
$Y=sc\varepsilon$, we have the simultaneous equations
\begin{align}
\label{a2}
a^2&=2(aX+bY), \\
b^2&+2(Xd-Yb)=1, \\
ba&=Y(a+d).
\end{align}
Solving for $a^2$ gives
\begin{align}
a^2 = 2[X^2+Y^2+\sqrt{(X^2+Y^2)^2+Y^2}].
\end{align}

Using that equilibrium value, the solution for $\sqrt\gamma A^T\bar x$ is
(for large time, so initial conditions may be ignored)
\begin{align}
\label{sdesol}
\sqrt\gamma A^T\bar x &= \gamma A^T \int_0^t \exp \left\{ \gamma[(\openone+
\varepsilon\Sigma)/2 \right. \nn & \!\!\!\!\!\!\!
\left.+(G^{-1}-\openone)AA^T](u-t)\right\}(G^{-1}-\openone)A\Delta I(u)du.
\nn &= \gamma \re \left( \A \int_0^t du e^{\gamma \Lambda (u-t)}\Delta I(u)
 \right),
\end{align}
where
\begin{align}
\A &= a-1 + i(a-2X)/\Delta, \\
2\Lambda &= a+ i\Delta,
\end{align}
with $\Delta^2/2 = \sqrt{(X^2+Y^2)^2+Y^2}-(X^2+Y^2)$.

Expanding $(\Delta I-\sqrt\gamma A^T\bar x)^2$ and taking the expectation value
gives
\begin{align}
\label{expect}
&\ex{(\Delta I-\sqrt\gamma A^T\bar x)^2} = \ex{\Delta I^2(t)} \nn &
- 2\gamma \re\left( \A \int_0^t du e^{\gamma\Lambda (u-t)} \ex{\Delta I(t)
\Delta I(u)} \right)\nn & + \frac{\gamma^2|\A|^2}2 \int_0^t du \int_0^t dv
e^{\gamma[\Lambda (u-t) + \Lambda^* (v-t)]} \ex{\Delta I(u)\Delta I(v)} \nn &
 +\frac {\gamma^2}2 \re \left( \A^2 \int_0^tdu\int_0^t dv
e^{\gamma\Lambda(u+v-2t)}\ex{\Delta I(u) \Delta I(v)} \right).
\end{align}
Determining $\langle I(u)I(v) \rangle$ gives
\begin{align}
&\langle I(u)I(v) \rangle = (1-S^2)\{\gamma \langle x(u) x(v) \rangle -
\sqrt\gamma [\langle x(u) \xi(v) \rangle \nn &\quad + \langle x(v) \xi(u)
\rangle] + \langle \xi(u) \xi(v) \rangle \} \nn & \quad
+S^2\{\gamma \langle y(u) y(v) \rangle + E^2 - \sqrt\gamma [
\langle y(u) \eta(v) \rangle \nn & \quad+ \langle y(v) \eta(u) \rangle] +
\langle \eta(u) \eta(v) \rangle \} ,
\end{align}
where $S=\sin(\Phi-\Theta)$. The solutions for $x$ and $y$ are
\begin{align}
x(t)&=\sqrt\gamma \int_0^t e^{\gamma(1+\varepsilon)(v-t)/2}\xi dv, \\
y(t)&=\sqrt\gamma \int_0^t e^{\gamma(1-\varepsilon)(v-t)/2}\eta dv.
\end{align}
Using this we obtain
\begin{align}
\langle I(u)I(v) \rangle &= \delta(u-v)+S^2 E^2+
S^2\frac{\gamma\varepsilon}{1-\varepsilon}e^{-\gamma(1-\varepsilon)|u-v|/2}
\nn & \quad +(S^2-1)\frac{\gamma\varepsilon}{1+\varepsilon}
e^{-\gamma(1+\varepsilon)|u-v|/2},
\end{align}
and
\begin{align}
\langle \Delta I(u)\Delta I(v) \rangle &= \delta(u-v)+(S-s)^2 E^2 \nn & \quad
+S^2\frac{\gamma\varepsilon}{1-\varepsilon}e^{-\gamma(1-\varepsilon)|u-v|/2}
\nn & \quad +(S^2-1)\frac{\gamma\varepsilon}{1+\varepsilon}
e^{-\gamma(1+\varepsilon)|u-v|/2}.
\end{align}

Using this result gives
\begin{align}
&\ex{(\Delta I-\sqrt\gamma A^T\bar x)^2}
= \delta(0)+ \gamma f(r,\Phi-\Theta,\theta-\Theta) \nn
& +[1-\re(\A/\Lambda)]^2(S-s)^2E^2, \label{71}
\end{align}
where $f(r,\Phi-\Theta,\theta-\Theta)$ is a function independent of $\gamma$ and
$E$; the explicit form will be given below. The term on the second line of
Eq.~\eqref{71} may be identified with the phase information obtained from the
mean field (as it is proportional to $E^2$), and is the most important
contribution. Expanding to second order, $(S-s)^2\approx\cos^2(\Phi-\Theta)
(\theta-\Theta)^2$, while
\begin{equation}
[1-\re(\A/\Lambda)]^2 = (c^2e^{-2r}+s^2e^{2r})^{-1}.
\end{equation}
In this expression we can replace $s$ with $S$, because we only need this term
to zeroth order in $\theta-\Theta$. We then have
\begin{align}
& \ex{(\Delta I-\sqrt\gamma A^T\bar x)^2} = \frac{E^2 \cos^2(\Phi-\Theta)}{\CS}
(\theta-\Theta)^2 \nn & + \gamma f(r,\Phi-\Theta,\theta-\Theta)+\delta(0)
+O((\theta-\Theta)^3),
\end{align}
where $\CS=(1-S^2)e^{-2r}+S^2e^{2r}$.

\begin{figure}
\centering
\includegraphics[width=0.45\textwidth]{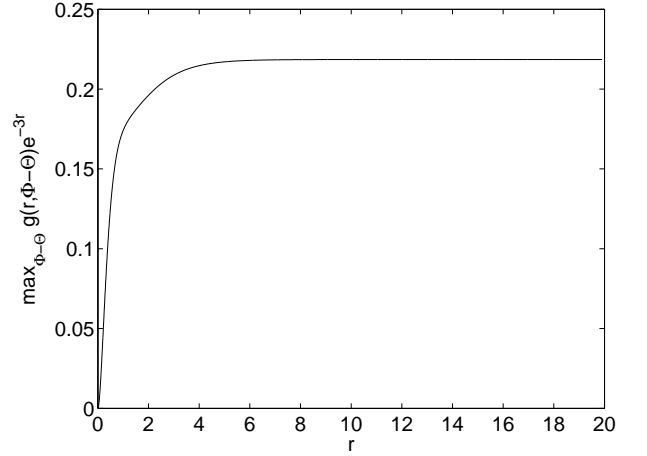}
\caption{The maximum of $g(r,\Phi-\Theta)$ divided by
$e^{3r}$ as a function of $r$.}
\label{fig1}
\end{figure}

The additional term $\gamma f(r,\Phi-\Theta,\theta-\Theta)$ gives the phase
information obtained from the noise. Its explicit form is
\begin{align}
& f(r,\Phi-\Theta,\theta-\Theta) = \re \left\{\frac{|\A|^2}{2 a}+\frac{\A^2}
{4\Lambda} \right. \nn & + S^2\frac{\varepsilon}{1-\varepsilon}\left[ 1+\frac
{{\A^2}/{2\Lambda}+{|\A|^2}/{a}-2\A}{\Lambda+(1-\varepsilon)/2}\right] \nn &
\left. +(S^2-1)\frac{\varepsilon}{1+\varepsilon}\left[ 1+\frac {{\A^2}/
{2\Lambda}+{|\A|^2}/{a}-2\A}{\Lambda+(1+\varepsilon)/2} \right] \right\}.
\end{align}
Expanding this function to second order in $\theta-\Theta$ gives
\begin{align}
f(r,\Phi-\Theta,\theta-\Theta) &= {\rm const.} + g(r,\Phi-\Theta)
(\theta-\Theta)^2 \nn & \quad +O((\theta-\Theta)^3),
\end{align}
where
\begin{align}
g(r,\Phi-\Theta) &= \frac{2\sinh^2(r/2)}{\CS a_{\Theta}}
\left[ (1+\cosh r)\Delta_\Theta^2 \right. \nn &\left.
+2\sin^2(\Phi-\Theta)\cosh r \left(1+\frac{\cosh r}{\sqrt{\CS}}\right)\right],
\end{align}
and $a_\Theta$ and $\Delta_\Theta$ are equal to $a$ and $\Delta$ for
$\theta=\Theta$. Numerics indicate that $g(r,\Phi-\Theta)\le e^{3r}/4$ (see
Fig.\ \ref{fig1}); therefore this term may be ignored provided $\gamma e^{3r}$
is small compared to $E^2/\CS$.

In Appendix \ref{ap:var}, the average over $\Theta$ is also taken. We take
$\Phi-\Theta$ to be a constant, rather than taking $\Phi$ to be independent of
$\Theta$. To obtain $\log P(\theta)$, we take the time integral of
$(\Delta I-\sqrt\gamma A^T\bar x)^2$ times $-1/2$. Therefore, for measurement
over a time interval $\Delta t$ we have
\begin{equation}
\label{result}
\frac 1{\Delta t\sigma^2} \approx \frac{E^2\cos^2(\Phi-\Theta)}
{\CS}+\gamma g(r,\Phi-\Theta).
\end{equation}
The first term gives the phase information due to the coherent amplitude,
whereas $\gamma g(r)$ gives the phase information due to the noise.
Note that $\lim_{r\to 0} g(r)=0$, so the phase information from the noise is
zero for coherent states, as we expect.

An essential point is the time required for the system to come to equilibrium.
The real part of the time constant $\gamma\Lambda$ is $a\gamma/2$. For small
$s$, we have $a/2\approx X$, so $\re(\Lambda)\approx 1/(e^r+1)$.
Therefore, the approximations made will only be accurate provided the
measurement is made over a time period that is long compared to
$(e^r+1)/\gamma$. Ignoring the constant 1, this is the $e^r/\gamma$ time scale
used in the previous section.

\section{Heterodyne measurements}
\label{sec:het}
We may derive similar results for the heterodyne case. The corresponding
equations are
\begin{align}
\dot x &= -x \gamma(1+\varepsilon)/2 +\sqrt{\gamma/2}(\xi^{(1)}+\xi^{(2)}) \\
\dot y &= -y \gamma(1-\varepsilon)/2 +\sqrt{\gamma/2}(\eta^{(1)}+\eta^{(2)}) \\
I_1 &= \cos(\Phi-\Theta)[\sqrt{\gamma/2}x-\xi^{(1)}]
\nn & \quad +\sin(\Phi-\Theta)[\sqrt{\gamma/2}(y+E')-\eta^{(1)}] \\
I_2 &= -\sin(\Phi-\Theta)[\sqrt{\gamma/2}x-\xi^{(2)}]
\nn & \quad +\cos(\Phi-\Theta)[\sqrt{\gamma/2}(y+E')-\eta^{(2)}].
\end{align}
The quantities $I_1$ and $I_2$ are the two Fourier components of the
photocurrents at the frequency at which the local oscillator is detuned from the
system \cite{WisMil93c}. Because of that detuning, $\Phi$ here is arbitrary.

To see that the results do not depend on the value of $\Phi$, note that
we may represent the measurement results by
the complex current $I = e^{i\Phi}(I_1+iI_2)$ given by
\begin{equation} \label{het1}
I = e^{i\Theta}\{\sqrt{\gamma/2}[x+i(y+E')]
-\nu_1-\nu_2\}.
\end{equation}
where $\nu_1=(\xi^{(1)}+i\eta^{(2)}+\xi^{(2)}+i\eta^{(1)})/2$
and $\nu_2=e^{2i(\Theta-\Phi)}(\xi^{(1)}+i\eta^{(2)}-\xi^{(2)}-i\eta^{(1)})/2$
are  independent complex Gaussian
random variables. Note also that
\begin{equation} \label{het2}
\dot x+i\dot y = -x\gamma(1+\varepsilon)/2-iy\gamma(1-\varepsilon)/2
+\sqrt{2\gamma}\nu_1.
\end{equation}

Following the derivation in the same way as above yields (see Appendix
\ref{app:het})
\begin{align}
& \frac{1}{\Delta t\sigma^2} = \frac{E^2}{1+e^{-2r}}
 + 2\gamma h(r),
\end{align}
where
\begin{equation}
h(r) = \cosh r - (\varepsilon^2+1)^{-1/2}.
\end{equation}
As in the adaptive case, the first term gives the phase information due to the
coherent amplitude, and $2\gamma h(r)$ gives the phase information due to the
noise. Again, $\lim_{r\to 0} h(r)=0$, so the phase information from the noise
is zero for coherent states. For large $r$, $h(r)$ scales as $e^{r}$; thus the
phase information from the noise will be negligible unless the contribution to
the photon flux due to the squeezing is dominant. In the derivation in
Appendix \ref{app:het} the $(e^r+1)/\gamma$ time constant again appears,
indicating that this is the appropriate time scale to observe squeezing in this
case also.

\section{Scaling revisited}
\label{sec:rev}
We now use these results to give a more detailed derivation of the scaling
for the phase variance. The derivations given in Secs.\ \ref{sec:bay} and
\ref{sec:het} are based on fixed system phase and feedback phase (for the
adaptive case). However, provided these phases vary by a small amount over
some time interval $\Delta t>(e^r+1)/\gamma$, these equations should still give
an accurate estimate of the phase information over this time interval.

First we consider the adaptive case. Provided the feedback phase is sufficiently
accurate, the $\cos$ in Eq.\ \eqref{result} may be replaced with 1, and
$\CS\approx e^{-2r}$. In addition, provided $\gamma e^r \ll E^2$ (which will be
true asymptotically unless the majority of the photon flux is from the
squeezing), the second term which gives the phase information from the noise may
be ignored. For continuous measurements, Eq.\ \eqref{result} gives the change in
the inverse variance. Therefore we have
\begin{equation}
\frac{1}{\Delta t}\Delta\left(\frac 1{\sigma^2}\right) \approx e^{2r} E^2.
\end{equation}
Taking into account the system phase varying as $\dot\Theta = \sqrt{\kappa}
\zeta$, we obtain the additional term $-\kappa/\sigma^4$ given in
Eq.\ \eqref{steady}.

We may then follow the derivation given in Sec.\ \ref{simple} to obtain the
scaling \eqref{scaling}. There are two assumptions used in this derivation which
we should justify. One assumption is that the phase information obtained from
the noise is negligible. For the scaling obtained in Sec.\ \ref{simple},
$E^2/\CS \sim \N e^{2r}$, whereas $\gamma e^{3r}<\N e^r$. Thus the assumption
that little phase information is obtained from the noise is reasonable.

The other assumption is that $\CS\sim e^{-2r}$. To check this condition, note
that $\sin^2(\Phi-\Theta)$ may be made as small as order $\sigma^2$ via
feedback. Therefore
\begin{equation}
\sin^2(\Phi-\Theta)e^{2r} \sim \left( \frac\kappa\N\right)^{5/8}
\left( \frac\N\kappa\right)^{1/4}
= \left( \frac\kappa\N\right)^{3/8}
\end{equation}
and
\begin{equation}
e^{-2r} \sim \left( \frac\kappa\N\right)^{1/4}>\sin^2(\Phi-\Theta)e^{2r}.
\end{equation}
Thus $\CS\sim e^{-2r}$, as required.

For the case of heterodyne measurements, the equivalent of Eq.\ \eqref{steady}
is
\begin{equation}
\label{hetsteady}
\frac{1}{\Delta t}\Delta\left(\frac 1{\sigma^2}\right) \approx -\frac{\kappa}
{\sigma^4} + \frac{E^2}{1+e^{-2r}}.
\end{equation}
Solving for $\sigma^2$ gives
\begin{equation}
\sigma^2 \approx \frac {\sqrt{1+e^{-2r}}}2 \sqrt{\frac\kappa{\N}}.
\end{equation}
Here we have used the approximation that $\N\approx E^2/4$, which simply means
that there is little photon flux from the squeezing. If there is no limit to the
squeezing, the asymptotic variance should be
\begin{equation}
\sigma^2 \approx \frac 12 \sqrt{\frac\kappa{\N}}.
\end{equation}

\section{Numerical results}
\label{sec:num}
As it is predicted that the phase information from the noise should be
negligible it is reasonable to consider a non-Bayesian method which does not
use phase information from the noise. The phase estimation method used for the
majority of the calculations was similar to that in Ref.\ \cite{cont}. The
quantities $A_t$ and $B_t$ were calculated as
\begin{align}
A_t &= \int_{-\infty}^t e^{\chi(u-t)}e^{i\Phi} I(u) du, \\
B_t &= -\int_{-\infty}^t e^{\chi(u-t)}e^{2i\Phi}.
\end{align}
Note that these quantities are not related to the $A$ and $B$ used in
Sec.\ \ref{sec:bay}. The $\chi$ used here is equivalent to that used in
Sec.\ \ref{sec:sqst}, in that the time scale over which previous measurement
results are used is $1/\chi$.

A good phase estimate may be obtained as
$\arg C_t$, where $C_t=A_t+\chi B_tA_t^*$ \cite{cont}. It was found that
poor results were obtained if $\arg C_t$ was used in the feedback, so
the feedback used was
\begin{equation}
\Phi(t) = \arg(C_t^{1-\delta}A_t^\delta)
\end{equation}
for a suitably chosen $\delta$.

In the calculations it is not necessary to independently vary $\N$ and $\kappa$.
We may scale the time by $\kappa$, so we obtain the dimensionless parameters
$\N/\kappa$, $\gamma/\kappa$ and $\chi/\kappa$. The other parameters which we
may vary are $r$ and $\delta$; these are already dimensionless. We may predict
scalings in terms of these dimensionless parameters:
\begin{align}
\sigma^2 &\sim ( \kappa/\N)^{5/8}, \quad
e^r \sim (\N/\kappa)^{1/8}, \nn
\frac\gamma\kappa &\sim (\N/\kappa)^{3/4}, \quad
\frac\chi\kappa \sim (\N/\kappa)^{5/8}.
\end{align}
The scaling of $\chi/\kappa$ is predicted from the fact that we use phase
information from a time interval $\sim \sigma^2/\kappa$.

Numerical calculations were performed for a range of values of $\N/\kappa$. For
each value six alternative measurement schemes were considered: \\
1. Adaptive measurements with arbitrary squeezing; \\
2. Adaptive measurements with limited squeezing; \\
3. Adaptive measurements on coherent states; \\
4. Heterodyne measurements with arbitrary squeezing; \\
5. Heterodyne measurements with limited squeezing; \\
6. Heterodyne measurements on coherent states. \\
In each case the appropriate parameters were adjusted to minimize the variance.
For all cases it was necessary to optimize over $\chi/\kappa$; for the adaptive
measurement it was also necessary to optimize over $\delta$. For the cases with
squeezing the values of $r$ and $\gamma/\kappa$ were also optimized over. In the
cases with limited squeezing there was the additional restriction that
$e^{2r}\le 2$. This limit corresponds to the typical maximum squeezing which may
be achieved in the laboratory.

\begin{figure}
\centering
\includegraphics[width=0.45\textwidth]{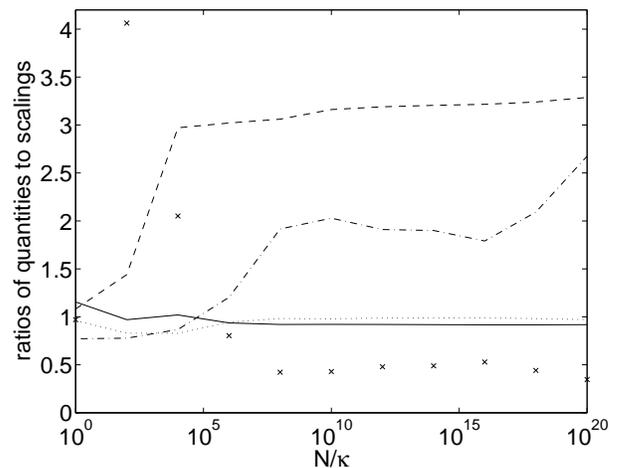}
\caption{The optimal values of various quantities as well as the variance for
adaptive measurements with arbitrary squeezing. The values of $\sigma^2/
(\N/\kappa)^{5/8}$ are shown as the solid line, $e^r/(\N/\kappa)^{1/8}$ is shown
as the dotted line, $(\gamma/\kappa)/(\N/\kappa)^{3/4}$ is shown as the dashed
line, $(\chi/\kappa)/(\N/\kappa)^{5/8}$ is shown as the dash-dotted line and
$\delta/(\N/\kappa)^{1/4}$ is shown as the crosses.} \label{fig2}
\end{figure}

Results were obtained by integrating the system over a time interval of
$130/\chi$, and determining the average phase variance for the data from time
$30/\chi$ to $130/\chi$. A total of $2^{10}$ independent integrations were
performed, so the total number of effectively independent samples, including
those from the different integrations as well as those from different times
within the integration, was approximately $10^5$.

The results for adaptive measurements on arbitrarily squeezed states are shown
in Fig.\ \ref{fig2}. In this figure the various quantities were divided by the
predicted scalings, in order to check that these scalings are correct. In each
case we find that this ratio is of order 1, so the predicted scalings are
correct. We also obtain the scaling for $\delta$ as
\begin{equation}
\delta \sim (\N/\kappa)^{1/4}.
\end{equation}
It does not appear to be possible to predict this scaling in the same way as the
scalings for the other parameters given here.

\begin{figure}
\centering
\includegraphics[width=0.45\textwidth]{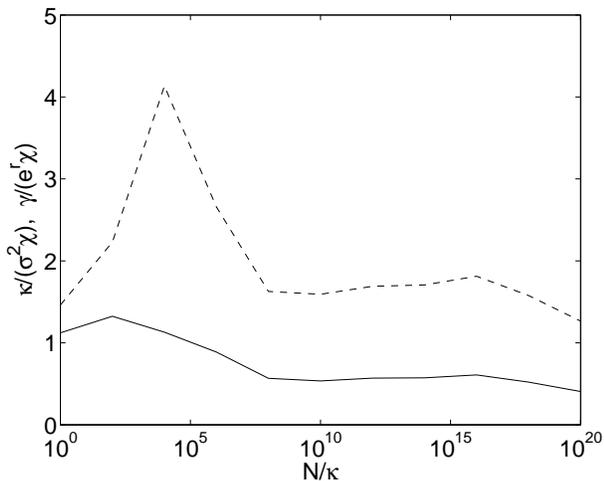}
\caption{The ratio of the value of $\kappa/\sigma^2$ to $\chi$ (continuous
line), and the ratio of $\gamma/e^r$ to $\chi$ (dashed line).} \label{fig3}
\end{figure}

The above theory predicts that $\chi$ should be
approximately equal to $\kappa/\sigma^2$ in order to minimize the variance.
Also, the time scale $1/\chi$ should be larger than the $e^r/\gamma$ time scale
required to obtain squeezing. To check these predictions, the ratio of
$\kappa/\sigma^2$ to $\chi$, as well as the ratio of $\gamma/e^r$ to $\chi$ are
shown in Fig.\ \ref{fig3}. From these results, the optimal value of $\chi$ is
near $\kappa/\sigma^2$, but there can be as much as a factor of two difference.
We find that $\gamma/e^r$ is larger than $\chi$, as predicted. However, the
difference is not great, and in some cases $\gamma/e^r$ is only slightly larger
than $\chi$.

\begin{figure}
\centering
\includegraphics[width=0.45\textwidth]{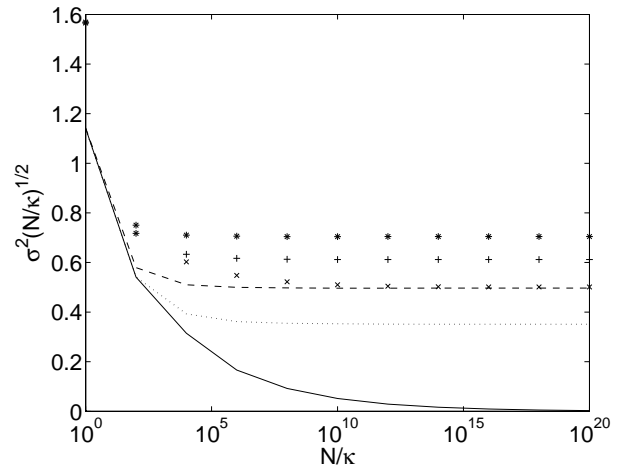}
\caption{The phase variance multiplied by $\sqrt{\N/\kappa}$. The results for
adaptive measurements on arbitrarily squeezed states, states limited to
$e^{2r}\le 2$, and coherent states are shown as the solid, dotted and dashed
lines, respectively. The results for heterodyne measurements on arbitrarily
squeezed states, states limited to $e^{2r}\le 2$, and coherent states are shown
as the crosses, plusses and asterisks, respectively.} \label{fig4}
\end{figure}

The phase variances for each of the six measurement schemes are shown in
Fig.\ \ref{fig4}. In this figure the variances have been multiplied by
$\sqrt{\N/\kappa}$ in order to more clearly show the scaling constants. In each
case except for adaptive measurements with arbitrarily squeezed states, it is
clear that the variance is scaling as $\sqrt{\kappa/\N}$.

It is straightforward to predict the asymptotic values of $\sigma^2
\sqrt{\N/\kappa}$ under each of the measurement schemes using the results given
in the preceding sections. These predictions are given in Table \ref{onlytable},
together with the asymptotic values estimated based on the numerical results.
In each case, the predicted and calculated asymptotic values are within 1\%.

\begin{table*}
\caption{Asymptotic values of $\sigma^2\sqrt{\N/\kappa}$ under each of the
measurement schemes. The numerically estimated values are given first, and the
analytic predictions are given in brackets.\label{onlytable}
\label{table}} \begin{tabular}{|c|c|c|} \hline
                                   & adaptive & heterodyne  \\ \hline
arbitrary squeezing                & 0 (0)                 & 0.501 ($1/2$)           \\
squeezing limited to $e^{2r}\le 2$ & 0.351 ($1/\sqrt 8$)   & 0.612 ($\sqrt{3/8}$)    \\
coherent states                    & 0.497 ($1/2$)         & 0.705 ($1/\sqrt{2}$)    \\
\hline
\end{tabular}
\end{table*}

\begin{figure}
\centering
\includegraphics[width=0.45\textwidth]{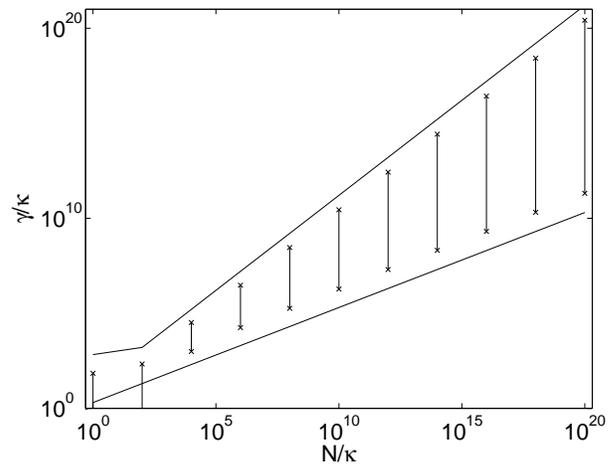}
\caption{The range of values of $\gamma/\kappa$ for which it is possible to
obtain phase variances within 10\% of the minimum values with limited squeezing
($e^{2r}\leq 2$) and adaptive measurements. The ranges are shown by the vertical
solid lines between the crosses, and the predicted upper and lower bounds
$2\N/(\kappa\sinh^2r)$ and $e^{2r}\sqrt{\N/\kappa}$ from Eq.~\eqref{14}
are shown by the diagonal
solid lines.}
\label{fig5}
\end{figure}

Another issue, raised at the end of Sec.~II~B,  is the range of possible
$\gamma$ values for which the benefit from squeezing may be
observed. In that section we derived the upper and lower bounds on $\gamma$ in
Eq.~\eqref{14}.  The range of values of $\gamma/\kappa$
such that the phase variance was still within 10\% of its minimum value was
calculated numerically for adaptive phase measurements, and is shown in
Fig.\ \ref{fig5}. (For the first two data points there was no lower bound found
numerically, because the variance was close to that for a coherent state.)

The scaling of the upper and lower bounds found numerically is the same as the
analytic bounds, though the range is slightly smaller. For the largest value of
$\N/\kappa$ tested, there is a range of nine orders of magnitude for $\gamma$.
This demonstrates that, even if it is not possible to adjust $\gamma$
experimentally, there will be a wide range of values for which a reduction in
the phase variance due to squeezing should be observed.

The last issue which we address in this section is that of using $\arg(C_t)$
rather than Bayesian phase estimates. To estimate the accuracy of the
$\arg(C_t)$ phase estimates, calculations were also performed with Bayesian
phase estimates. The system was integrated over a time period of $10^3/\chi$,
and the phase variance was estimated by averaging from time $30/\chi$. The
functions $P(\theta)$, $\bar x(\theta)$ and $G(\theta)$ were estimated by
calculating them at 2000 values of $\theta$. There is a complication when we
take account of the variation in the system phase. If the Bayesian analysis is
performed exactly, the distribution is no longer Gaussian in $\vec x$ for given
$\theta$. It is not feasible to perform the calculation for the full
distribution in $\vec x$ and $\theta$, so the distribution was approximated by a
Gaussian in $\vec x$. The effect of the varying system phase on the distribution
was approximated by simply adding a spread to the phase distribution
$P(\theta)$.

\begin{figure}
\centering
\includegraphics[width=0.45\textwidth]{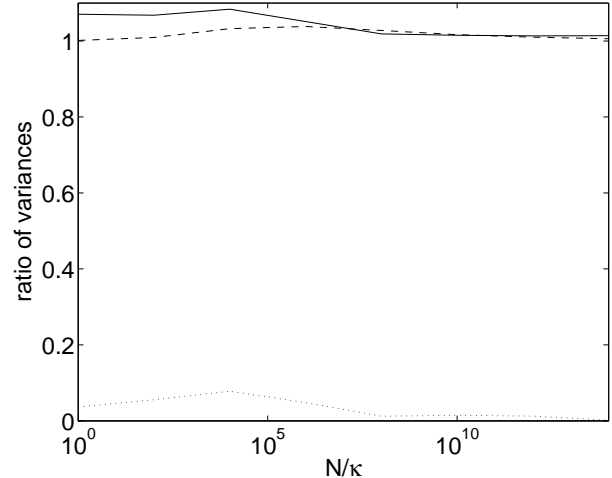}
\caption{The ratio of the phase variance obtained via $\arg(C_t)$ to the phase
variance obtained via the Bayesian method (solid line). The estimated ratio,
based on the difference being due to the phase information from the noise and
using the maximum value of $g(r,\Phi-\Theta)$, is shown as the dashed line.
The mean-square difference between the two phase estimates, as a ratio to the
phase variance for Bayesian estimates, is shown as the dotted line.}
\label{fig6}
\end{figure}

For each set of data, the phase was estimated both via the Bayesian method and
as $\arg(C_t)$. This allows accurate comparison of the relative variance. The
ratio of the variance for the $\arg(C_t)$ phase estimates to that for the
Bayesian phase estimates is shown in Fig.\ \ref{fig6}. There is only a small
difference between the two variances; on average about 4\%, and no more than
9\%. In fact, the phase estimates obtained via $\arg(C_t)$ and the Bayesian
method are quite close. The mean-square difference between these phase estimates
is only about 3\% of the total phase variance. This means that the qualitative
results obtained for the $\arg(C_t)$ phase estimates should also hold when
Bayesian phase estimates are used. The only difference is a reduction in the
variance of a few percent.

The difference between the variances is likely due to the fact that the Bayesian
estimate uses phase information from the noise, whereas the $\arg(C_t)$ estimate
does not. Using the result \eqref{result} we may predict the ratio of the
variances. This estimate is also shown in Fig.\ \ref{fig6}. There is not exact
agreement with the numerically obtained ratio, but the estimate is close,
particularly for the larger values of $\N/\kappa$.

\section{Conclusions}
\label{sec:conc} We have undertaken a thorough analysis of adaptive estimation
of a continuously varying phase for narrowband squeezed beams. The problem is
characterized by photon flux $\N$, phase diffusion rate $\kappa$, squeezing
bandwidth $\gamma$, and degree of squeezing $e^{2r}$ (i.e.\ depth of squeezing
of $1-e^{-2r}$). If it is possible to achieve arbitrarily high squeezing, then
adaptive phase measurements should give a variance scaling as
$(\N/\kappa)^{-5/8}$, an improvement over the $(\N/\kappa)^{-1/2}$ scaling for
coherent beams. This variance is higher than that suggested by the broadband
analysis in Ref.\ \cite{cont}, which had a scaling of $(\N/\kappa)^{-2/3}$. The
reason for this difference is that a time scale of $e^r/\gamma$ is required
before squeezing is observed. This means that the phase information must be
obtained from a longer time interval than in the broadband case, and the system
phase varies by a larger amount over this time interval. In the broadband case
the limiting factor was the accuracy of the feedback.

It is somewhat surprising that the time scale required to observe squeezing is
$e^r/\gamma$, rather than $1/\gamma$, which is what we would expect since
$\gamma$ is the squeezing bandwidth. The rate $\gamma/e^r$ is the bandwidth of
the antisqueezed quadrature, but it is only through the full Bayesian analysis
that one finds that this is the important rate. The difference of $e^r$ is
crucial, because the scaling of the phase variance would be $(\N/\kappa)^{-2/3}$
(as suggested by the broadband analysis) without it. The numerical calculations
verify that the scaling is $(\N/\kappa)^{-5/8}$ rather than
$(\N/\kappa)^{-2/3}$, demonstrating that the factor of $e^r$ is correct.

We performed a Bayesian analysis of the phase, for both adaptive and nonadaptive
(heterodyne) measurement schemes. In each case the analysis yields a
variance-reducing term proportional to $E^2$, which may be interpreted as the
phase information from the coherent amplitude, as well as a term proportional to
$\gamma$, which corresponds to phase information from the squeezed noise. Except
in cases where the dominant contribution to the photon flux is from the
squeezing, the phase information from the noise is predicted to be negligible.
This means that it is possible to obtain accurate phase estimates using a
simplified method rather than the full Bayesian estimate.

In the experimentally realistic case where there is limited squeezing, the
reduction in the phase variance is approximately $e^r$. Experimentally it is not
possible to produce beams with arbitrary values of $\gamma$, though a limited
amount of control is possible \cite{perscomm}. For larger values of $\N/\kappa$,
the variance is insensitive to the exact value of $\gamma$, and similar
variances are obtained for a wide range of values of $\gamma$. Therefore this
issue is not expected to be a problem experimentally. A more complete analysis
of the experimental feasibility of adaptive phase estimation on continuous
squeezed beams will be given in a future work.

\appendix
\section{Expected variance}
\label{ap:var} The standard measure for phase uncertainty, used in most of the
Refs.~\cite{single,cont,pope} is the Holevo variance \cite{Hol84}. It is defined
as $|\ex{e^{i\Delta\theta}}|^{-2}-1$, where $\Delta\theta$ is the error in the
phase estimate; that is, the estimated phase minus the actual system phase. An
alternative definition which avoids phase estimates with systematic error is
$\re\ex{e^{i\Delta\theta}}^{-2}-1$ \cite{thesis}. The Holevo variance under this
definition can be determined from \cite{thesis}
\begin{equation}
\ex{e^{i\Delta\theta}} = \frac 1{2\pi}\int dI \left| \int P(I|\theta)
e^{i\theta} d \theta \right|.
\end{equation}
Here $I$ is used to indicate the entire measurement record. This may be
alternatively written as
\begin{align}
\ex{e^{i\Delta\theta}} &= \int P(I) dI \left| \int P(\theta|I)
e^{i\theta} d \theta \right| \nn
&= \frac 1{2\pi} \int d\Theta \int P(I|\Theta) dI \left| \int P(\theta|I)
e^{i\theta} d \theta \right| .
\end{align}

The measurement record $I$ can be expressed as a function of $\Theta$ as well as
a record of random fluctuations $\nu$ which are independent of $\Theta$.
Therefore we may give the expression for $\ex{e^{i\Delta\theta}}$ as
\begin{equation}
\ex{e^{i\Delta\theta}} = \frac 1{2\pi} \int d\Theta \int P(\nu) d\nu
\left| \int P(\theta|\nu,\Theta)e^{i\theta} d \theta \right| .
\end{equation}
Provided the measurement gives small variance, the only values of $\theta$
and $\Theta$ for which $P(\nu)P(\theta|\nu,\Theta)$ is not negligible are those
for which $\theta$ is close to $\Theta$. Thus it is reasonable to expand
$P(\theta|\nu,\Theta)$ as a series about $\Theta$:
\begin{equation}
\ex{e^{i\Delta\theta}} \approx \frac 1{2\pi} \int d\Theta
\int P(\nu) d\nu \left| \int e^{a+b(\theta-\Theta)-c(\theta-\Theta)^2+i\theta}
 d \theta \right|
\end{equation}
where $a$, $b$ and $c$ are functions of $\nu$ and $\Theta$. Simplifying gives
\begin{align}
\ex{e^{i\Delta\theta}} &\approx \frac 1{2\pi} \int d\Theta
\int P(\nu) e^{-1/(4c)} d\nu \nn
& = \frac 1{2\pi} \int d\Theta \ex{e^{-1/(4c)}}_{\Theta}.
\end{align}
Provided the variance of $c$ is small, it is possible to use the approximation
\begin{equation}
\ex{e^{i\Delta\theta}} \approx 1-\frac 1{8\pi} \int d\Theta
(\ex{c}_{\Theta})^{-1}.
\end{equation}
If $\ex{c}_{\Theta}$ is independent of $\Theta$, $\ex{e^{i\Delta\theta}}
\approx 1-1/(4\ex{c})$, so the variance is approximately $1/2\ex{c}$.

\section{Heterodyne derivation}
\label{app:het}
The discretized version of the equations for heterodyne detection, (\ref{het1})
and (\ref{het2}) is
\begin{align}
&\Delta x+i\Delta y = - \gamma \Delta t [x(1+\varepsilon)/2+iy(1-\varepsilon)/2]
+ \! \sqrt{2\gamma\Delta t}\tilde\nu_1\nn
&I = e^{i\Theta}[ \sqrt{\gamma/2} (x+iy+iE')-(\tilde\nu_1+\tilde\nu_2)
/\sqrt{\Delta t}].
\end{align}
Here the $\tilde\nu_1$, $\tilde\nu_2$ are complex Gaussian random variables with
$\ex{\tilde{\nu}_j} = 0 =  \ex{\tilde{\nu}_i\tilde{\nu}_j}$ but
$\ex{\tilde{\nu}_i^*\tilde{\nu}_j} = \delta_{ij}$.

At each time step we update the probability $P(x,y,\theta)$,
ignoring the constant factor $1/P(I)$, by
\begin{equation}
P(x,y,\theta) \to P(x,y,\theta)\times P(I|x,y,\theta).
\end{equation}
The probability $P(I|x,y,\theta)$ is given by
\begin{align}
P(I|x,y,\theta)&\propto \exp \left( -\frac{\Delta t}2 \left|
I-e^{i\theta}\sqrt{\gamma/2}(x+iy+iE') \right|^2\right) \nn &=
\exp\left(-\frac{\gamma\Delta t}4 [(x-I_x)^2+(y-I_y)^2]\right),
\end{align}
where $I_x+iI_y=\sqrt{2/\gamma}Ie^{-i\theta}-iE'$.
At all times the probability $P(x,y,\theta)$ is given by
\begin{equation}
P(x,y,\theta) = P(\theta)P(x|\theta)P(y|\theta),
\end{equation}
where $P(x|\theta)$ and $P(y|\theta)$ are normal distributions for $x$ and $y$.
$P(\theta)$ gives the correct probability for $\theta$ averaging over
$x$ and $y$.

We therefore have
\begin{align}
\frac 1{\sigma_x^2} &\mapsto \frac 1{\sigma_x^2}+\frac{\gamma\Delta t}2, \qquad
\frac 1{\sigma_y^2} \mapsto \frac 1{\sigma_y^2}+\frac{\gamma\Delta t}2, \nn
\bar x &\mapsto (\bar x+\sigma_x^2I_x\gamma\Delta t/2)/(1+\sigma_x^2\gamma\Delta
 t/2),\nn
\bar y &\mapsto (\bar y+\sigma_y^2I_y\gamma\Delta t/2)/(1+\sigma_y^2\gamma\Delta
 t/2).
\end{align}
In the infinitesimal limit,
\begin{align}
d\sigma_x^2 = -\sigma_x^4\gamma dt/2,& \qquad
d\sigma_y^2 = -\sigma_y^4\gamma dt/2, \nn
d\bar x = \sigma_x^2(I_x-\bar x)\gamma dt/2,& \qquad
d\bar y = \sigma_y^2(I_y-\bar y)\gamma dt/2.
\end{align}
For the change in the probability distribution for $\theta$ we have,
in the infinitesimal limit
\begin{equation}
P^{(2)}(\theta) \propto P(\theta)\exp\left\{ -\frac {\gamma dt}4 [(I_x-\bar x)^2
+(I_y-\bar y)^2]\right\}.
\end{equation}

Now we take account of the increments in $x$, $y$ and $\theta$. We have the
restriction that
\begin{align}
\tilde\nu_1+\tilde\nu_2 = \sqrt{\Delta t} [\sqrt{\gamma/2}(x+iy+iE')-
Ie^{-i\theta}].
\end{align}
The mean of $\tilde\nu_1$ is therefore half of the RHS, and the second-moment is
1/2, rather than 1. Therefore, the difference equation may be expressed as
\begin{equation}
\Delta x+i\Delta y = -[(x-iy)\varepsilon+I_x+iI_y] \gamma \Delta t/2
 +\tilde\nu\sqrt{\gamma\Delta t},
\end{equation}
where $\tilde\nu$ is a complex Gaussian random variable. Considering the
deterministic part of the increment first, and taking the infinitesimal limit
\begin{align}
d\sigma_x^2 = -\sigma_x^2\varepsilon \gamma dt,& \qquad
d\sigma_y^2 = \sigma_y^2\varepsilon \gamma dt, \nn
d\bar x = -(I_x+\varepsilon \bar x)\gamma dt/2,& \qquad
d\bar y = -(I_y-\varepsilon\bar y)\gamma dt/2.
\end{align}
To take account of the stochastic increments, we simply add the appropriate
variances:
\begin{align}
d\sigma_x^2 = \gamma dt/2, \qquad
d\sigma_y^2 = \gamma dt/2.
\end{align}
Overall we have the increments
\begin{align}
d\sigma_x^2 &= [1-2\varepsilon\sigma_x^2-\sigma_x^4]\gamma dt/2 , \nn
d\sigma_y^2 &= [1+2\varepsilon\sigma_y^2-\sigma_y^4]\gamma dt/2 .
\end{align}
Therefore the steady state values are
\begin{equation}
\sigma_x^2 = \sqrt{\varepsilon^2+1}-\varepsilon, \qquad
\sigma_y^2 = \sqrt{\varepsilon^2+1}+\varepsilon.
\end{equation}

The total increments in $\bar x$ and $\bar y$ are
\begin{align}
d\bar x &= -[\bar x (\sigma_x^2+\varepsilon) + (1-\sigma_x^2)I_x] \gamma dt/2, \nn
d\bar y &= -[\bar y (\sigma_y^2-\varepsilon) + (1-\sigma_y^2)I_y] \gamma dt/2.
\end{align}
The solutions are
\begin{align}
\bar x &= (\sigma_x^2-1)(\gamma/2)\int_0^t I_x e^{(\gamma/2)\sqrt{\varepsilon^2+1}(v-t)}dv, \nn
\bar y &= (\sigma_y^2-1)(\gamma/2)\int_0^t I_y e^{(\gamma/2)\sqrt{\varepsilon^2+1}(v-t)}dv.
\end{align}
Now $\langle I \rangle = i e^{\Theta} \sqrt{\gamma/2} E'$, so
$\langle I_x+iI_y \rangle = i (e^{i(\Theta-\theta)}-1)E'$, and
\begin{align}
\langle \bar x \rangle &= \sin(\theta-\Theta) E' \frac{\sigma_x^2-1}{\sqrt{\varepsilon^2+1}}, \nn
\langle \bar y \rangle &= [\cos(\theta-\Theta)-1] E' \frac{\sigma_y^2-1}{\sqrt{\varepsilon^2+1}}.
\end{align}

The solutions for $x$ and $y$ are
\begin{align}
x(t) &= \sqrt{2\gamma} \int_0^t e^{\gamma(1+\varepsilon)(u-t)/2}\re(\nu_1) du, \nn
y(t) &= \sqrt{2\gamma} \int_0^t e^{\gamma(1-\varepsilon)(u-t)/2}\im(\nu_1) du.
\end{align}
In terms of these quantities,
\begin{widetext}
\begin{align}
I_x(t) &= x(t) \cos(\theta-\Theta)+(y(t)+E')\sin(\theta-\Theta)
-\sqrt{2/\gamma} \{\re(\nu_1+\nu_2)\cos(\theta-\Theta)+\im(\nu_1+\nu_2)\sin(\theta-\Theta)\} \nn
I_y(t) &= -x(t) \sin(\theta-\Theta)+(y(t)+E')\cos(\theta-\Theta)-E'
-\sqrt{2/\gamma}\{\im(\nu_1+\nu_2)\cos(\theta-\Theta)-\re(\nu_1+\nu_2)\sin(\theta-\Theta)\}.
\end{align}
Therefore
\begin{align}
\ex{I_x(u)I_x(v)} &= \{\ex{x(u)x(v)}-\sqrt{2/\gamma}\ex{x(u) \re[\nu_1(v)]
+ x(v) \re[\nu_1(u)]}\}\cos^2(\theta-\Theta) \nn
& \quad + \{\ex{y(u)y(v)}-\sqrt{2/\gamma}\ex{y(u) \im[\nu_1(v)] + y(v) \im[\nu_1(u)] }
+(E')^2\}\sin^2(\theta-\Theta) +(2/\gamma)\delta(u-v) \nn
\ex{I_y(u)I_y(v)} &= \{\ex{x(u)x(v)}-\sqrt{2/\gamma}\ex{x(u) \re[\nu_1(v)]
+ x(v) \re[\nu_1(u)]}\}\sin^2(\theta-\Theta) \nn
& \quad + \{\ex{y(u)y(v)}-\sqrt{2/\gamma}\ex{y(u) \im[\nu_1(v)] + y(v) \im[\nu_1(u)] }\}
\cos^2(\theta-\Theta) \nn
& \quad +(E')^2[\cos(\theta-\Theta)-1]^2 +(2/\gamma)\delta(u-v).
\end{align}
Substituting gives
\begin{align}
\ex{I_x(u)I_x(v)} &= e^{-r} e^{-\gamma(1+\varepsilon)|u-v|/2}\cos^2(\theta-\Theta)
+ \{e^r e^{-\gamma(1-\varepsilon)|u-v|/2}
+(E')^2\}\sin^2(\theta-\Theta)+(2/\gamma)\delta(u-v) \nn
\ex{I_y(u)I_y(v)} &= e^{-r} e^{-\gamma(1+\varepsilon)|u-v|/2}\sin^2(\theta-\Theta)
+ e^r e^{-\gamma(1-\varepsilon)|u-v|/2}\cos^2(\theta-\Theta) \nn
& \quad +(E')^2[\cos(\theta-\Theta)-1]^2
+(2/\gamma)\delta(u-v).
\end{align}

Using this result gives
\begin{align}
\langle (I_x-\bar x)^2+(I_y-\bar y)^2\rangle
&= {\rm const.}+(E')^2\left(1-\frac{\sigma_x^2-1}{\sqrt{\varepsilon^2+1}}\right)^2
\sin^2(\theta-\Theta)+(E')^2\left(1-\frac{\sigma_y^2-1}{\sqrt{\varepsilon^2+1}}\right)^2
[\cos(\theta-\Theta)-1]^2 \nn & \quad +
\left(\frac{e^r}{\sigma_x^2+1}-\frac{e^{-r}}{\sigma_y^2+1}\right)
\left[\frac{(\sigma_x^2-1)^2-(\sigma_y^2-1)^2}{\sqrt{\varepsilon^2+1}}
-2(\sigma_x^2-\sigma_y^2)\right]\sin^2(\theta-\Theta).
\end{align}
Here the terms that do not depend on $\theta$ or $\Theta$ have been collected
into the constant. Expanding to second order in $(\theta-\Theta)$ and using the
equilibrium values of $\sigma_x^2$ and $\sigma_y^2$ gives
\begin{align}
\frac{\gamma dt}2\langle (I_x-\bar x)^2+(I_y-\bar y)^2\rangle &= {\rm const.}+
\frac{E^2}{1+e^{-2r}}(\theta-\Theta)^2
 +2\gamma dt \left[ \cosh r -(\varepsilon^2+1)^{-1/2} \right](\theta-\Theta)^2.
\end{align}
\end{widetext}

\acknowledgments
This research has been supported by the Australian Research Council. The
authors acknowledge valuable discussions with Elanor Huntington, Hideo Mabuchi,
and Tim Ralph.

\end{document}